\documentclass{ws-rv9x6}
\pdfoutput=1
\usepackage{subfigure}   
\usepackage{ws-rv-thm}   
\usepackage[square]{ws-rv-van}   
\makeindex
\usepackage{graphicx}
\usepackage{wrapfig}
\usepackage{lscape}
\usepackage{rotating}
\usepackage{epstopdf}
\usepackage{caption} 

\begin{document}

\chapter[Deep Learning From Four Vectors]{Deep Learning From Four Vectors}

\author{Pierre Baldi}
\address{Department of Computer Science, UC Irvine, Irvine CA, USA 92627}
\author{Peter Sadowski}
\address{Department of Computer Science, University of Hawaii, XYZ}
\author{Daniel Whiteson}
\address{Department of Physics \& Astronomy, UC Irvine, Irvine CA, USA 92627}

\begin{abstract}
An early example of the ability of deep networks to improve the statistical power of data collected in particle physics experiments was the demonstration that such networks operating on lists of particle momenta (four-vectors) could outperform shallow networks using features engineered with domain knowledge. A benchmark case is described, with extensions to parameterized networks. A discussion of data handling and architecture is presented, as well as a description of how to incorporate physics knowledge into the network architecture.
\end{abstract}
\body

\tableofcontents

\clearpage

\section{Introduction: Pre-Deep Learning State-of-the-Art}

The goal of the statistical analysis of particle physics data is to infer bounds on parameters of physical theories, such as the masses of particles or their rate of production in specific interactions.  These statistical tasks, which involve classification, hypothesis testing, regression, and
goodness-of-fit testing, require a statistical likelihood model $p(x |
\theta)$ which describes the probability of observing experimental data $x$ for specific values of the
parameters $\theta$ of a physical theory.  

Unfortunately, the statistical likelihood model can almost never be expressed analytically, due to the complex nature of the relationship between the theoretical parameters $\theta$ and the high-dimensional ($10^2- 10^8)$) data $x$. Instead, statistical models are typically estimated from samples generated using  Monte Carlo methods~\cite{Alwall:2011uj, Agostinelli:2002hh}, whose  computational expense limits the dimensionality of the feature space to $\mathcal{O}(10^0)$.  In this context, it becomes vital to reduce the dimensionality of the data, and many initial applications~\cite{Abramowicz:1995zi,Abazov2001282,ABREU1992383} of machine learning to particle physics focused on development of classifiers, which offered powerful ways to perform this dimensional reduction and produce a single feature which summarizes much of the available information relevant to the statistical question.

Early applications of machine learning in physics \cite{vazquez1992improving,baldi2021deep}
were largely limited to shallow machine learning methods including boosted decision trees and artificial neural networks with a single hidden layer. This was primarily for historical reasons, including broad unawareness of the power of deep neural networks, misguided thinking and publications about local minima or vanishing gradients, as well as concerns over the problem of interpreting their output. 
In addition, it was well known that neural networks with a single hidden layer have universal approximation properties \cite{hornik1989multilayer}, although highly non-linear functions may require an intractable number of hidden nodes. The combination of these circumstances led physicists to focus on shallow rather than deep machine learning methods, and shallow classifiers rapidly proliferated in physics
(see \cite{vazquez1992improving} for an early application of neural networks). The approach successfully boosted the performance of many statistical analyses by allowing physicists to employ multiple observables, and was commonly referred to as ``multi-variate analysis". Such applications reduced the dimensionality of the feature space to one or two, allowing for estimation of the statistical models needed for inference tasks.  


While dimensional reduction nearly always involves some loss of information, it does not necessarily reduce relevant information, as the statistical task usually only requires a subset of the information. The Neyman-Pearson lemma shows that the optimal decision boundary for a hypothesis test between two statistical models (in {\it any} dimension feature space) can be determined by knowledge of their ratios; the full models are not needed.  However, it was long suspected that shallow networks fell short of capturing all of the relevant information contained in the feature space.  While it is not possible to directly estimate the absolute optimal performance without building an optimal classifier, it is possible to demonstrate that a given network is not optimal by finding a more powerful example. 

A common experience in pre-deep-learning particle physics was to perform exhaustive feature engineering to simplify the task for the shallow classifier.  A shallow network on four vectors\footnote{Four vectors in momentum space are a generalization of three-dimensional momentum, including the total energy $E$, as $(E,\bar{p})$ and are typically used in particle physics to specify a particle's momentum and total energy.}, for example, would be compared to a shallow network with features built using domain knowledge.  Such domain-specific expert features are generally non-linear functions of four vectors that capture physical insights about the data. Almost invariably, the expert features would boost the performance, despite adding no unique information, demonstrating that the shallow classifiers had  failed to discover these non-linear strategies on their own.

This feature-search approach is labor-intensive and not necessarily optimal; a robust machine learning method would obviate the need for this additional step and capture all of the available classification power directly from the raw data. Thus, the stage was set for deep learning.

\section{Application of Deep Learning to Four Vectors}

In this section we describe a benchmark classification task\cite{Baldi:2014kfa,Baldi:2014pta} that exemplifies a common use case for machine learning in particle physics: discrimination between signal and background processes. This example demonstrates a common failure mode of shallow networks on four vectors, which exhibit reduced performance compared to networks that use features engineered with domain knowledge.

\subsection*{Benchmark Case for Higgs Bosons}

\begin{figure}[ht]
\centering

\subfigure[ ]{
\includegraphics[width=2in]{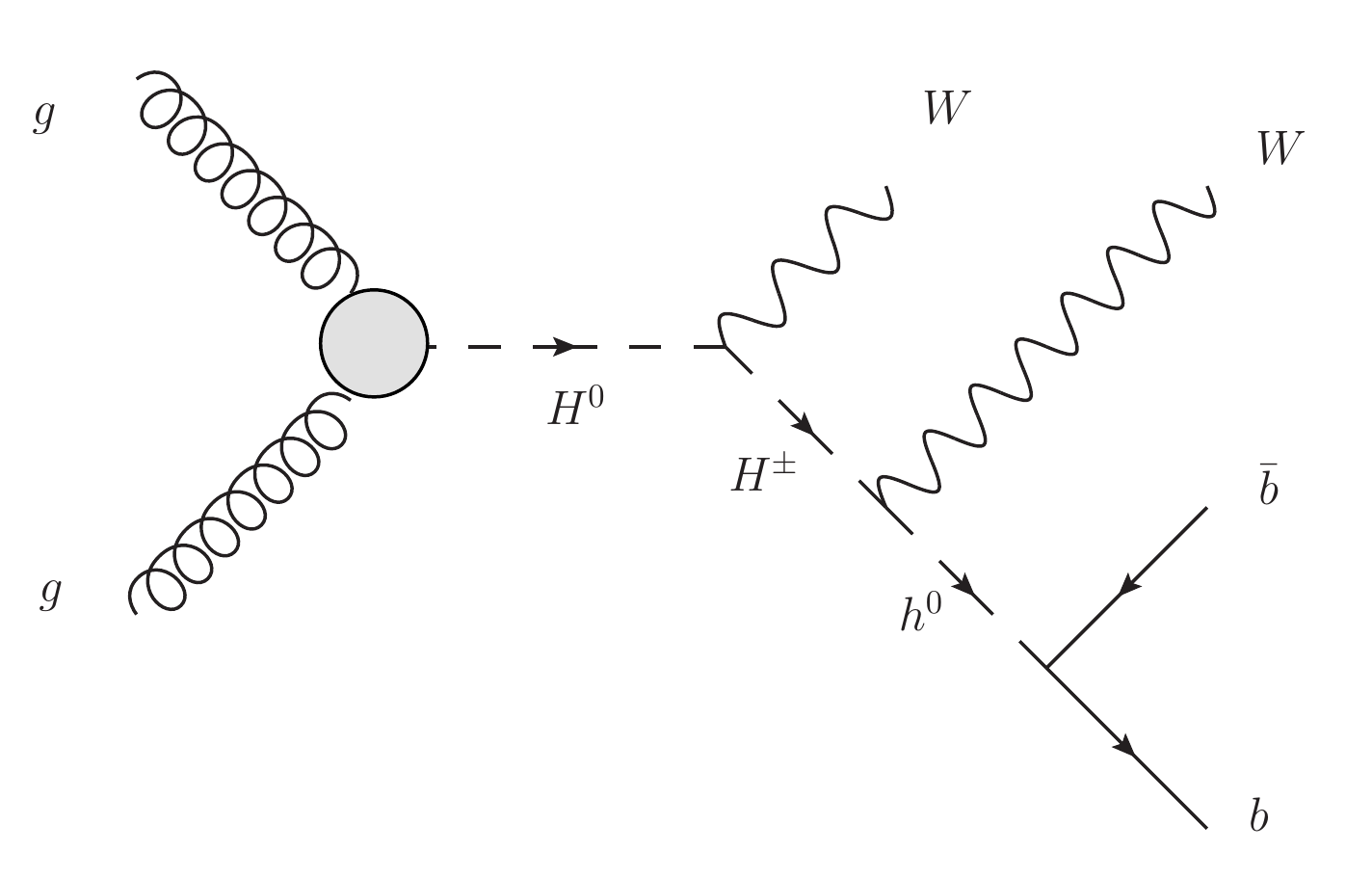}
\label{fig:ttbb_a}
}
\subfigure[ ]{
\includegraphics[width=2in]{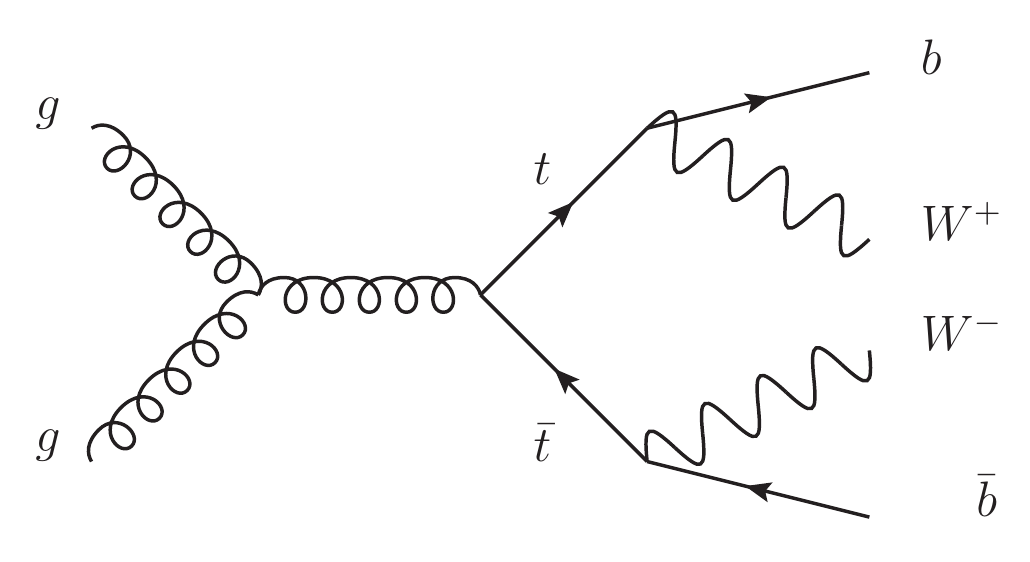}
\label{fig:ttbb_b}
}
\caption{Diagrams describing (a) the signal process involving new exotic Higgs bosons $H^0$ and $H^\pm$; and (b) the background process involving top-quarks ($t$). In both cases, the resulting particles are two $W$ bosons and two $b$-quarks.}
\label{fig:ttbb}
\end{figure}

A typical classification task in particle physics distinguishes between a signal process, where new particles are produced, and one or more background processes, which mimic the nature and number of particles observed, but can be distinguished by their kinematics. An example~\cite{Baldi:2014kfa} examined by experiments at the LHC is the production of a heavy electrically-neutral Higgs boson ($gg\rightarrow H^0$), which decays to a heavy electrically-charged Higgs boson ($H^\pm$) and a $W$ boson\cite{cdfwwbb,atlaswwbb}. The $H^\pm$ boson subsequently decays to a second $W$ boson and the light Higgs boson, $h^0$ observed by the ATLAS~\cite{atlashiggs}  and CMS~\cite{cmshiggs} experiments. The light Higgs boson decays predominantly to a pair of bottom quarks, giving the process:
\begin{equation}
gg\rightarrow  H^0\rightarrow W^{\mp}H^{\pm}\rightarrow W^\mp W^\pm h^0\rightarrow
  W^\mp W^\pm b\bar{b}, 
\end{equation}

\noindent
which leads to $W^\mp W^\pm b\bar{b}$ shown in Fig.~\ref{fig:ttbb}. For the benchmark case here, $m_{H^0}=425$ GeV and $m_{H^\pm}=325$ GeV have been assumed.

The background process mimics this signal but without the Higgs boson intermediate state. It produces a pair of top quarks, each of which decay to $Wb$, also giving $W^\mp W^\pm b\bar{b}$.  

Both processes yield the same set of observed particles: one charged lepton, four jets (two of which have $b$-tags) and missing transverse momentum.  Together, the twenty-one individual momentum components of these particles comprise our {\it low-level feature set}.

\begin{figure}[ht]
\centering
\subfigure[ ]{
\includegraphics[width=1.5in]{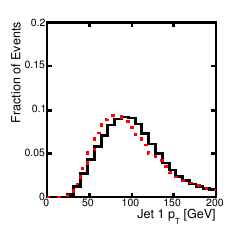}
\label{fig:llvar_a}}
\subfigure[ ]{
\includegraphics[width=1.5in]{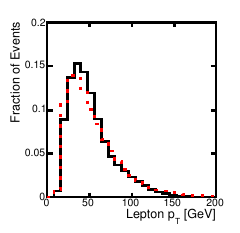}
\label{fig:llvar_e}}
\caption{Distributions of example low-level input features for Higgs benchmark. 
Shown are the simulated signal (black solid) and background (red dotted) marginal distributions of transverse momenta ($p_{\rm T}$) of (a) the most energetic jet and (b) the lepton. }
\label{fig:llvar}
\end{figure}

\begin{figure}
\centering
\subfigure[ ]{
\includegraphics[width=1.5in]{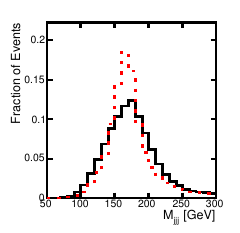}
\label{fig:hlvar_b}}
\subfigure[ ]{
\includegraphics[width=1.5in]{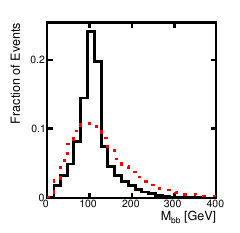}
\label{fig:hlvar_e}}
\caption{Distributions of example high-level features for Higgs benchmark. Shown are the simulated signal (black) and background (red) events for marginal distributions of (a) the three-jet invariant mass and (b) the bottom-quark and anti-bottom-quark pair.}
\label{fig:hlvar}
\end{figure}

The low-level features show some differences between the signal and background processes --- Fig.~\ref{fig:llvar} shows the marginal distributions for two of these kinematic features. However, we see even larger differences in higher-level features constructed using domain knowledge of the different intermediate states.  As the difference in the two hypotheses lies mostly in the existence of new intermediate Higgs boson states, it is possible to distinguish between the two hypotheses by attempting to identify whether the intermediate state existed by reconstructing its characteristic invariant mass. In the signal hypothesis we expect peaks in $m_{\ell\nu}$, $m_{jj}$, $m_{b\bar{b}}$, $m_{Wb\bar{b}}$, $m_{WWb\bar{b}}$, while the background should peak in $m_{j\ell\nu}$ and $m_{jjj}$. Figure~\ref{fig:hlvar} shows the difference in distributions for two of these high level variables. 

\subsection*{Performance}

\begin{figure}[h]
\centering
\subfigure[ ]{
\includegraphics[width=2in]{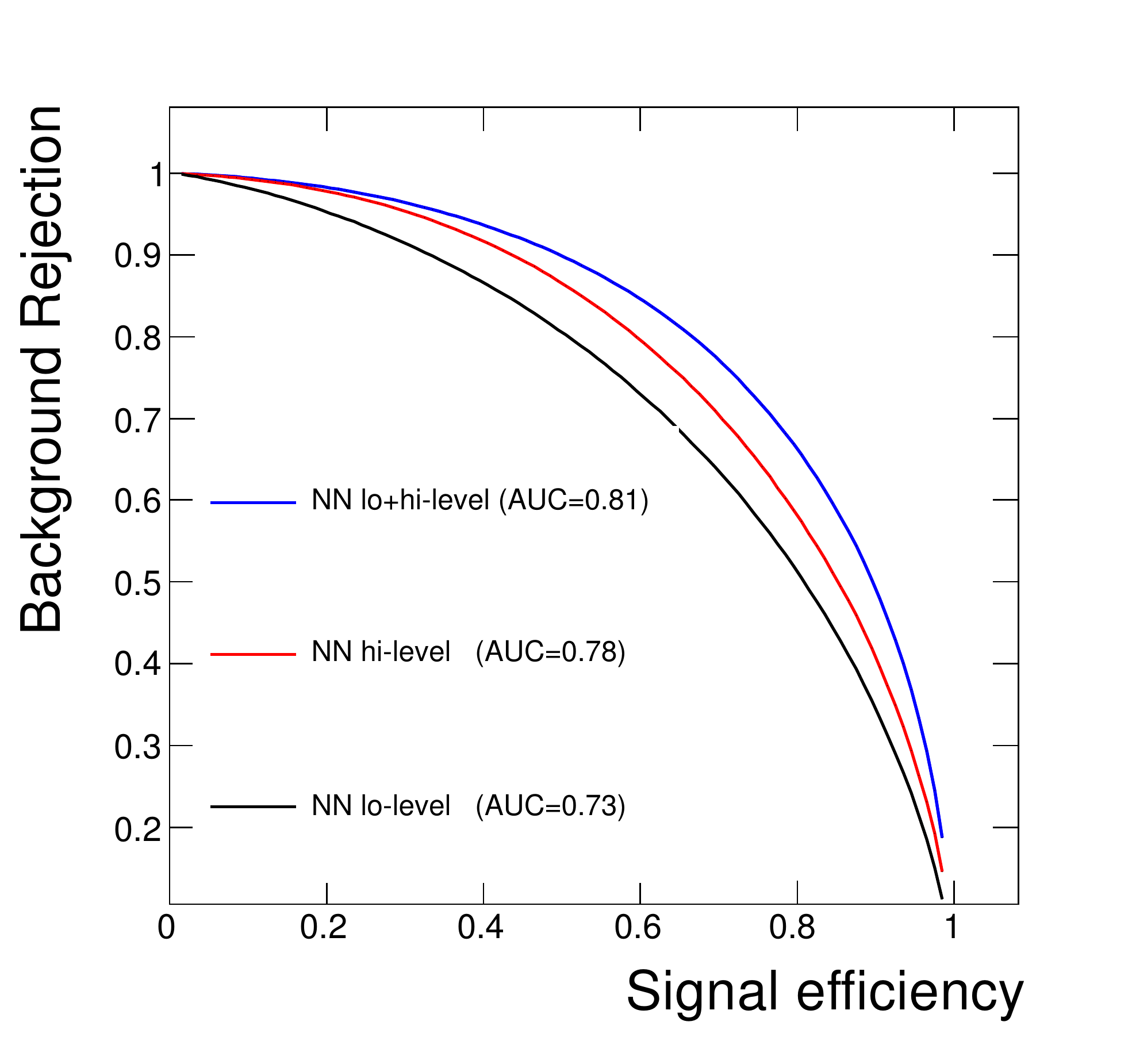}
\label{fig:auc_a}}
\subfigure[ ]{
\includegraphics[width=2in]{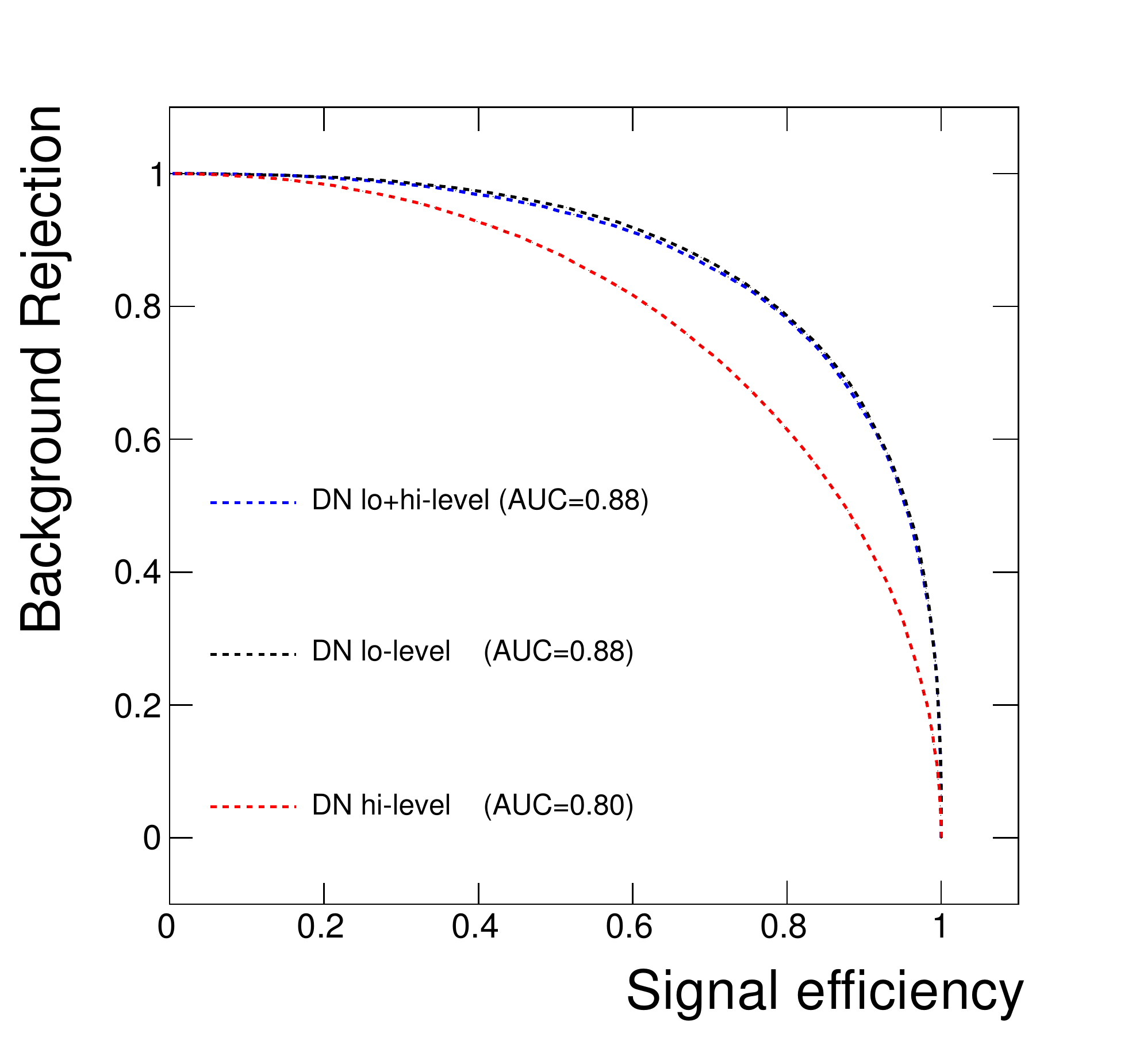}
\label{fig:auc_b}}
\caption{Background rejection versus signal efficiency on the Higgs benchmark for (a) shallow neural networks (NN) and (b) deep neural networks (DN). Curves are plotted for models trained using the low-level features only (black), the high-level features only (red), and the complete set of features (blue).}
\label{fig:efficiency}
\end{figure}

Deep neural networks (DNN) were compared to shallow neural networks (NN) and boosted decision trees (BDT) on three different subsets of input features: the low-level features only, the high-level features only, and both. Performance on the test set was measured in terms of Area Under the ROC curve (AUC) and discovery significance (Table~\ref{tab:auc}), as well as signal efficiency and background rejection; see Fig.~\ref{fig:efficiency}.

\begin{table}[h]
\centering
\begin{tabular}{llll}
\hline\hline
 & \multicolumn{3}{c}{AUC}\\
Technique & Low-level & High-level & Complete \\
\hline
BDT & 0.73 ($0.01$) & 0.78  ($0.01$)& 0.81  ($0.01$) \\
NN 	& $0.733$ ($0.007$)	& $0.777$ ($0.001$) &  $0.816$ ($0.004$) \\
DNN	& $0.880$ ($0.001$)	& $0.800$ ($<0.001$) &  $0.885$ ($0.002$) \\
\hline\hline
 & \multicolumn{3}{c}{Discovery significance}\\
Technique & Low-level & High-level & Complete \\
\hline
NN 	& $2.5\sigma$ & $3.1\sigma$ & $3.7\sigma$\\
DNN	& $4.9\sigma$ & $3.6\sigma$ & $5.0\sigma$\\
\hline\hline
\end{tabular}
\caption{ Comparison of the performance of several learning techniques: boosted decision trees (BDT), shallow neural networks (NN), and deep neural networks (DNN) for three sets of input features: low-level features, high-level features and the complete set of features. Each neural network was trained five times with different random initializations. The table displays the mean Area Under ROC Curve (AUC) of the signal-rejection curve in Fig.~\ref{fig:efficiency}, with statistical uncertainty measured in cross-validation shown in parentheses. Below is shown the expected significance of a discovery (in units of Gaussian $\sigma$) for 100 signal events and $1000\pm50$ background events.}
\label{tab:auc}
\end{table}

The shallow neural networks and BDTs trained with the high-level features perform significantly better than those trained on only the low-level features, demonstrating the importance of feature engineering in shallow machine learning models. However, training all three methods with \textit{only} the high-level features leads to lower performance than training with the complete set of features, indicating that the low-level features contain additional information that is not being captured by these engineered features. Only the deep learning approach shows nearly equal performance using the low-level features and the complete features. This suggests that it is automatically discovering high-level abstractions similar to those captured by the hand-engineered features, obviating the need for laborious feature engineering.

\subsection*{Discussion}

It is widely accepted in experimental high-energy physics that machine learning is a powerful approach boosting statistical power in exotic particle searches.  Until the advent of deep learning, physicists reluctantly accepted the limitations of the shallow machine learning classifiers --- laboriously constructing non-linear feature combinations to help guide shallow networks and BDTs. This benchmark study shows that advances in deep learning can lift these limitations by \textit{automatically} discovering powerful feature combinations directly from low-level features. Similar conclusions were reached in other benchmark cases, such as in searches for $t\bar{t}h$ production\cite{Santos:2016kno}.

\section{Parameterized Networks}

The deep learning approach described above solves a simple signal versus background classification task for a hypothesized particle with a particular mass. But the mass of a hypothesized particle is generally unknown. In practice, a hypothesized particle will have a \textit{range} of possible masses, each of which would produce a different type of signal in the data. The classification tasks at different masses are closely related, but distinct. Physicists need a way to evaluate this \textit{class} of signal hypotheses against the null (background) hypothesis.


A naive way to address this problem is to perform a finite number of individual comparisons. For each of $K$ possible mass values, a hypothesized particle with that mass is simulated using Monte Carlo to produce a data set, and a machine learning model is trained to discriminate between it and the background. Methods from statistics can be used to account for the problem of multiple-hypothesis testing (e.g. the Bonferroni correction~\cite{Dunn1961}). However, this naive approach is too conservative when the data distributions from different mass values are related. In practice, the distribution is expected to vary smoothly with the continuous parameter of the particle --- that is, one expects similar mass values to result in similar distributions of observation data. This suggests the use of machine learning to model the relationship between the mass parameter and the data distribution.

A machine learning solution to this problem is to train a \textit{single} classifier to perform particle searches for an entire range of possible mass values\cite{Cranmer:2015bka,Baldi:2016fzo}. This is done by extending the list of input features to include one or more additional parameters that describe the larger scope of the problem such as a new particle's mass. The approach can be applied to any classification model; however, neural networks provide a smooth interpolation in this new parameter space, while tree-based classifiers may not. A single parameterized network can replace a set of individual networks trained for specific cases, as well as smoothly interpolate to cases where it has not been trained. In the case of a search for a hypothetical new particle, this greatly simplifies the task -- by requiring only one network -- as well as making the results more powerful -- by allowing them to be interpolated between specific values. In addition, they may outperform isolated networks by generalizing from the full parameter-dependent dataset. In this section we describe the use of parameterized neural networks and provide a realistic example. For a real application, see Refs.\cite{sirunyan2018search,ATLAS:2018ibz,Sirunyan:2019nfw, 
Sirunyan:2019arl}. 

\subsection{Parameterized Network Structure \& Training}

\begin{figure}
\centering
\includegraphics[width=0.8\textwidth]{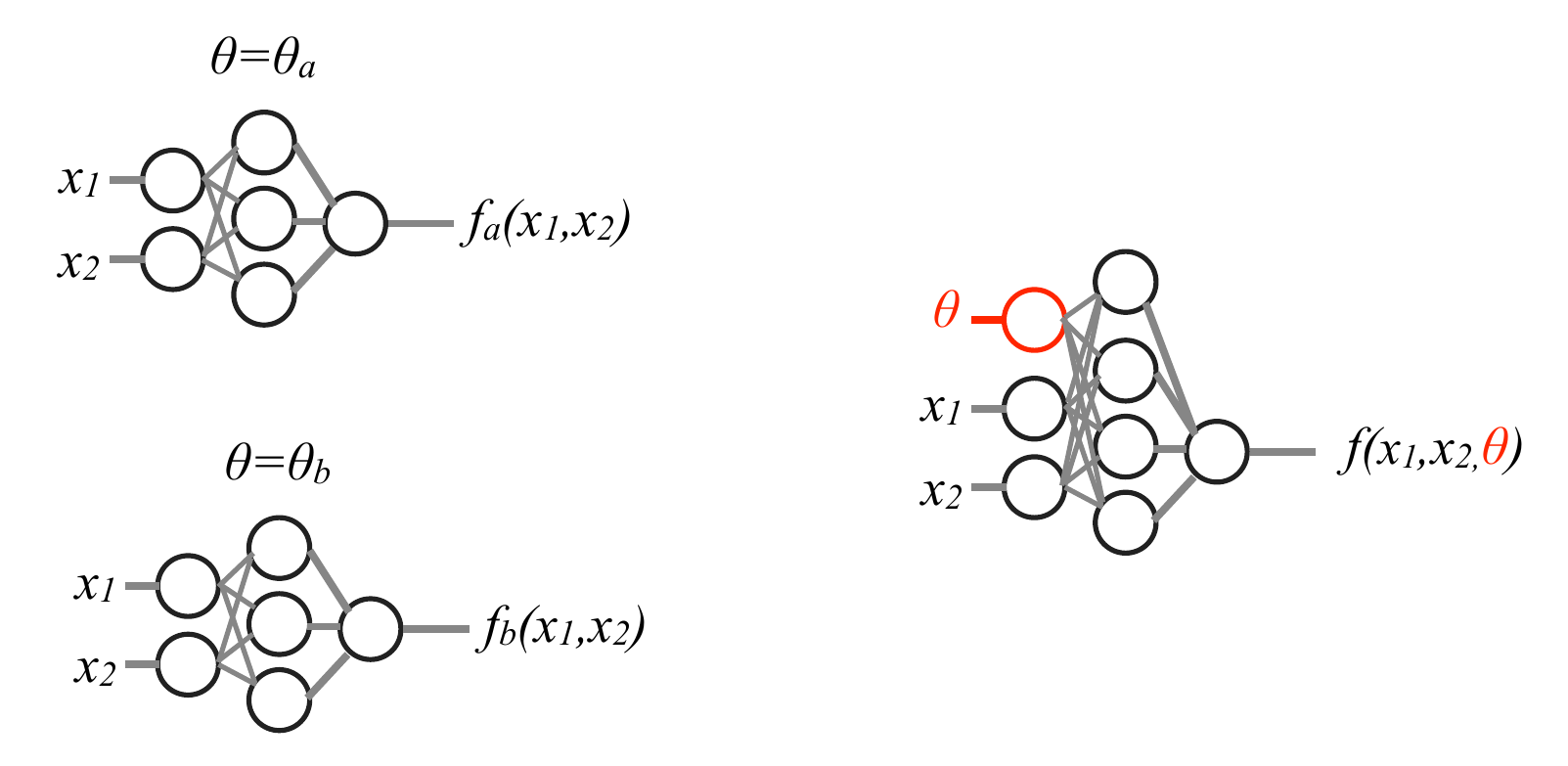}
\caption{Traditional neural networks (left) with input features $(x_1,x_2)$ are trained with examples from fixed values of some latent parameter $\theta=\theta_a,\theta_b$. Neither network performs optimally for intermediate values of $\theta$. A \textit{parameterized} network (right) is trained with input features $(x_1,x_2)$ as well as the input parameter $\theta$; such a network is trained with examples at several values of the parameter $\theta$ and interpolates for intermediate values. At test time, the user provides $\theta$.}
\label{fig:param_nn_diagram}
\end{figure}

A standard neural network takes as input a vector of features, $\bar{x}$, and computes a function of these features, $f(\bar{x})$. Parameterized networks address the case where the task is part of a larger context, described by one or more parameters, $\bar{\theta}$, by computing a function of both inputs: $f(\bar{x},\bar{\theta})$. Thus, a parameterized neural network makes different predictions for input $\bar{x}$ for different contexts $\bar{\theta}$; see Fig.~\ref{fig:param_nn_diagram}.  Unlike other networks, a parameterized network {\it requires} a value of $\bar{\theta}$ to perform inference on input $\bar{x}$, as the network output is a function of both.

Parameterized neural networks require some additional considerations during training. Each training example for such a parameterized network has the form $(\bar{x}, \bar{\theta}, y)_i$, where $y$ is the target output. However, in classification problems $\bar{\theta}$ may not be meaningful for a particular target class. For example, the mass of a new particle is not meaningful for the background training examples. To avoid divulging information about $y$, one must randomly assign values~\cite{Baldi:2016fzo} to $\bar\theta$ according to the same distribution used for the signal class.

Another issue is that the distribution of $\bar{\theta}$ in the training set represents a prior distribution that influences the final model, and should be specified carefully. Traditionally, $\bar{\theta}$ is determined by the hypothesis and fixed during detector simulations, producing samples from a conditional distribution $p(\bar{x} | \bar{\theta}, y)$; in parameterized neural networks, the training data would typically be generated  by first sampling $\bar{\theta}$ from a class-specific prior $p(\bar{\theta}| y)$ then $\bar{x}$ from $p(\bar{x} | \bar{\theta}, y)$. The robustness of the resulting parameterized classifier to the distribution of $\bar\theta$ in the training sample will depend on the physics encoded in the distributions $p(\bar{x} | \bar{\theta}, y)$ and how much they change with $\bar\theta$. The prior $p(\bar{\theta}| y)$ should be chosen carefully and should be considered when interpreting results, just as one would carefully consider a fixed $\bar{\theta}$ when building and evaluating a traditional classifier. In the studies presented below, the training data consists of equal sized samples for a few discrete values of $\bar{\theta}$ --- the conditional distribution $p(\bar{x} | \bar{\theta}, y)$ varies smoothly enough in $\bar{\theta}$ that this reasonably approximates a uniform prior over $\bar{\theta}$.

\subsection{Physical Example}

\begin{figure}
\centering
\subfigure[ ]{
\includegraphics[width=0.4\textwidth]{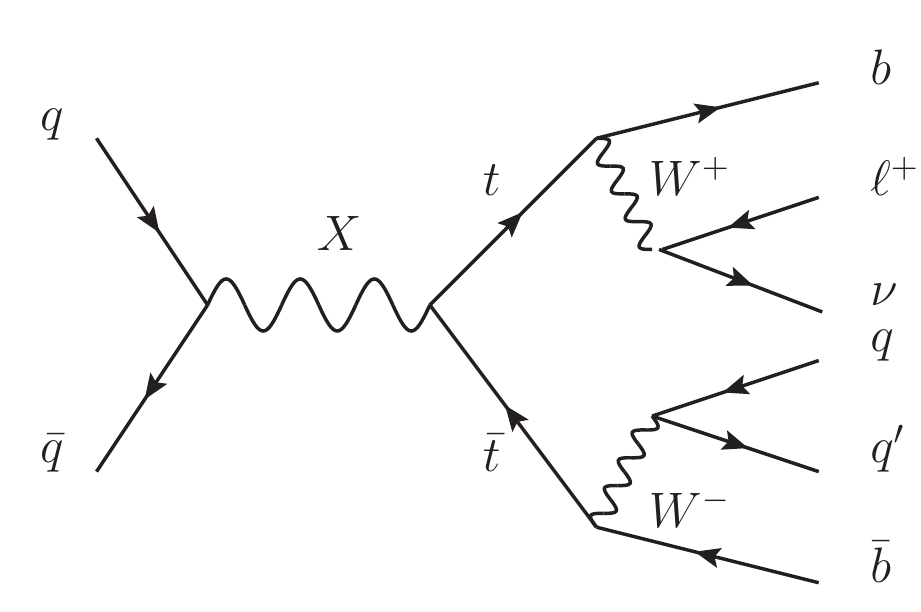}
}
\subfigure[ ]{
\includegraphics[width=0.4\textwidth]{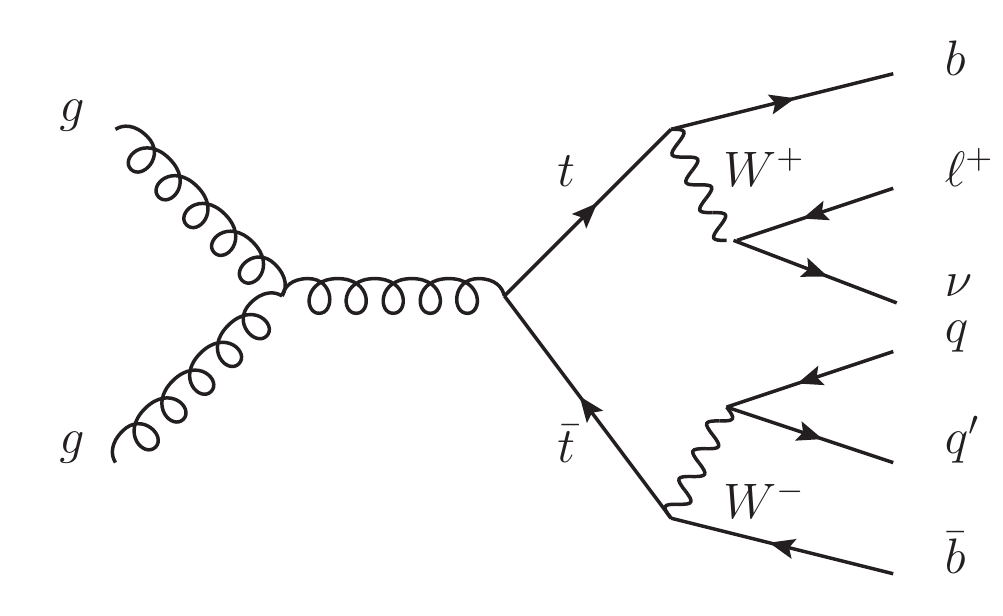}
}
\caption{Feynman diagrams showing (a) the production and decay of the hypothetical particle $X\rightarrow t\bar{t}$, as well as (b) the dominant standard model background process of top quark pair production. In both cases, the $t\bar{t}$ pair decay to a single charged lepton ($\ell$), a neutrino ($\nu$) and several quarks ($q,b$).}
\label{fig:param_feynman}
\end{figure}

Parameterized networks address a common problem in searches for new particles of unknown mass, and we provide an illustrative example from Ref.~\cite{Baldi:2016fzo}. Consider the search for a new particle $X$ which decays to $t\bar{t}$, examining the most powerful decay mode in which $t\bar{t}\rightarrow W^+bW^-\bar{b}\rightarrow qq'b\ell\nu \bar{b}$. The dominant background is standard model $t\bar{t}$ production, which is identical in final state but distinct in kinematics due to the lack of an intermediate resonance. Figure~\ref{fig:param_feynman} shows diagrams for the signal and background processes.

\begin{figure}
\centering
\includegraphics[width=0.4\textwidth]{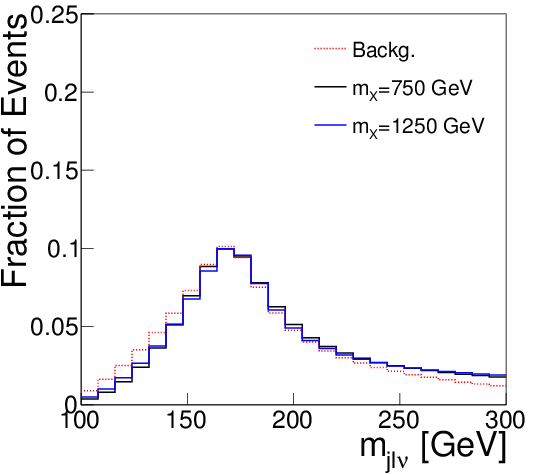}
\includegraphics[width=0.4\textwidth]{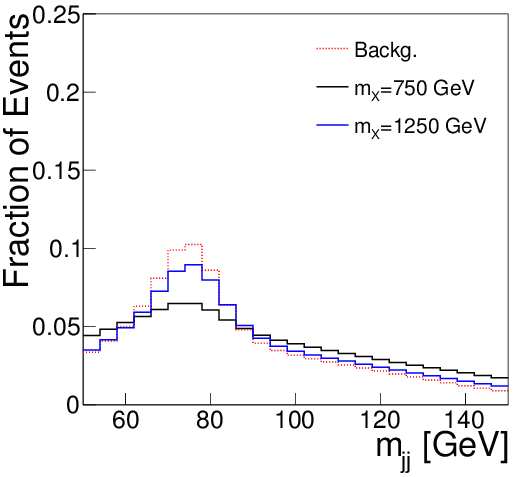}\\
\includegraphics[width=0.4\textwidth]{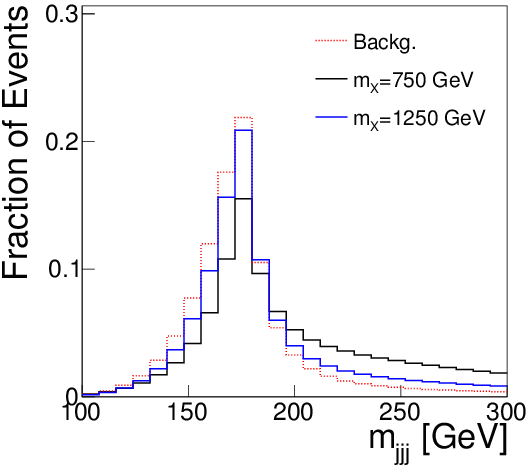}
\includegraphics[width=0.4\textwidth]{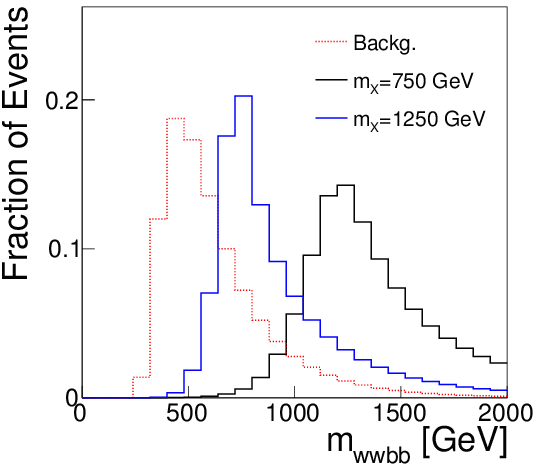}
\caption{Distributions~\cite{Baldi:2016fzo} of high-level event features for the decay of $X\rightarrow t\bar{t}$ with two choices of $m_X$ as well as the dominant background process; see text for definitions.}
\label{fig:param_hlvar}
\end{figure}

The set of event-level features include 21 low-level kinematic features resulting from reconstruction algorithms and 5 high-level features which incorporate physics domain knowledge. The distributions of four high-level features are shown in Fig.~\ref{fig:param_hlvar} to illustrate the differences between the signal distribution at different values of the particle mass, $m_X$, and the background distribution.  

In order to test how well a parameterized neural network generalizes to new parameter values, an experiment compared the performance of a fixed neural network architecture trained on three different training data sets with different distributions of $m_X$, and tested on a data with $m_X=1000$~GeV. The three different training data sets contained signal samples with different mass distributions: (1) $m_X=1000$~GeV only; (2) $m_X=500,750,1000, 1250, 1500$ GeV; and (3) $m_X=500,750, 1250, 1500$~GeV (no $m_X=1000$~GeV). In each case, the training set contains 7M examples, the test set contains 1M, and approximately the same number of training and testing examples are used per mass point. On each data set, the same neural network architecture was trained, containing five 500-dimensional ReLU layers followed by a logistic output unit for binary classification. Parameters were initialized from a Gaussian distribution with mean zero and width 0.1, and updated using stochastic gradient descent with mini-batches of size 100 and 0.5 momentum. The learning rate was initialized to 0.1 and decayed by a factor of 0.89 every epoch. Training was stopped after 200 epochs.

\begin{figure}
\centering
\includegraphics[width=0.7\textwidth]{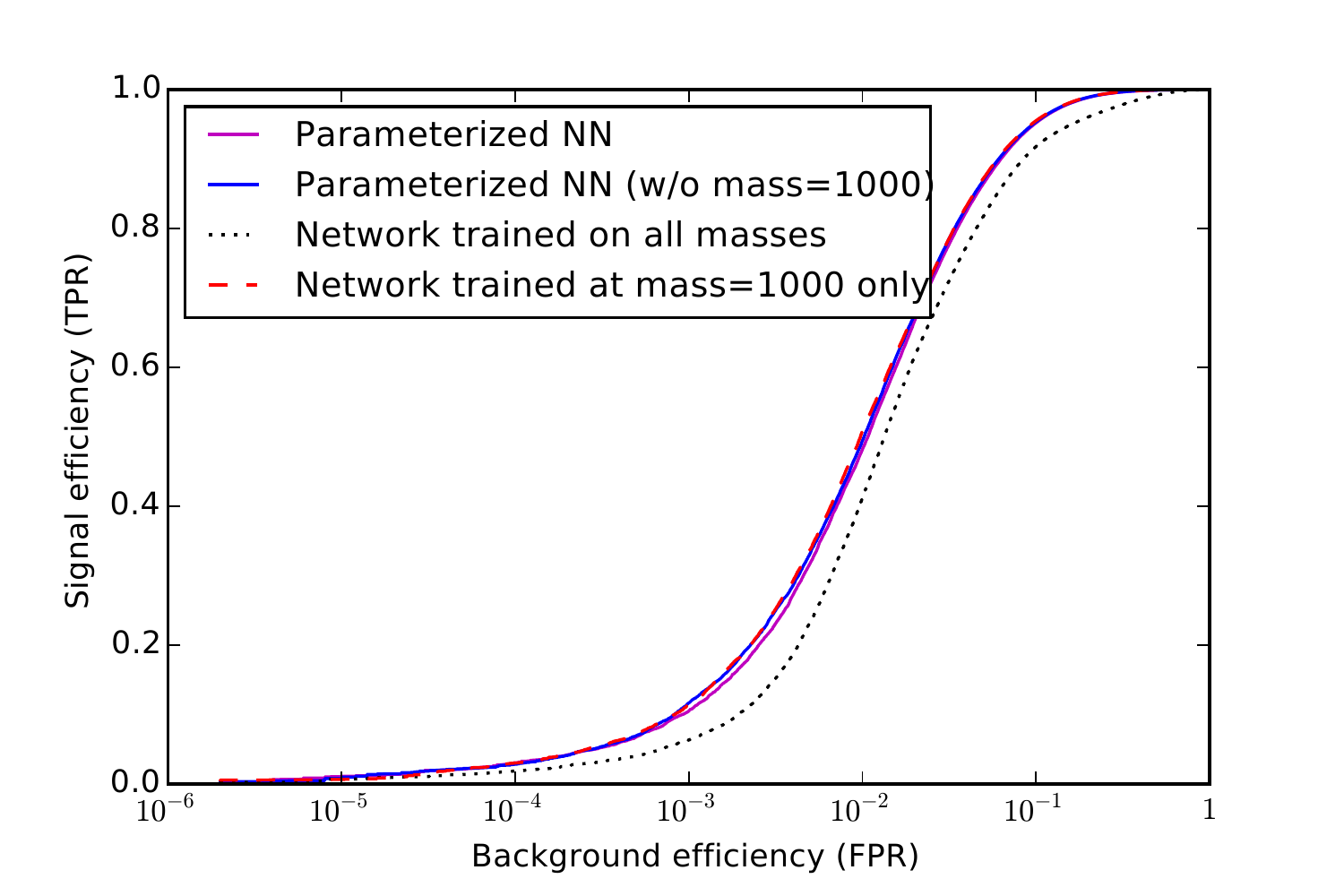}
\caption{Performance comparison~\cite{Baldi:2016fzo} of signal-to-background discrimination for four classes of networks on a test sample with $m_X=1000$ GeV. A parameterized network trained on all masses $m_X=500,750,1000,1250, 1500$ (magenta) performs just as well as a traditional network trained with only $m_X=1000$ GeV (red). A second parameterized network trained with only $m_x=500,750,1250,1500$ is forced to interpolate the solution at $m_X =1000$ GeV (blue), but performs equally well. However, a traditional non-parameterized network trained with all the mass points (black) shows a reduced performance. The results are indistinguishable for cases where the networks use only low-level features (shown) or low-level as well as high-level features (not shown, but identical).}
\label{fig:param_roc}
\end{figure}

The results show that the parameterized network not only matches the performance of a network trained on a single mass value, but is able to generalize to mass values it has never seen before. Figure~\ref{fig:param_roc} shows that the parameterized network trained on $m_X=500,750,1000, 1250, 1500$ GeV matches the performance of the fixed network trained on $m_X=1000$ only. In the third data set, $m_X=1000$ samples are removed from the training set so that the network must interpolate its solution, but the performance is unchanged, demonstrating that the parameterized network is able to generalize even in this high-dimensional example.

We note, however, that while the ability of the parameterized network was demonstrated in this case, and we expect this ability to generalize due to networks excellent performance in interpolation tasks, one cannot claim to predict similar quality of interpolation for an arbitrary task. Performance in a specific task would require a dedicated study.


\begin{figure}
\centering
\includegraphics[width=0.7\textwidth]{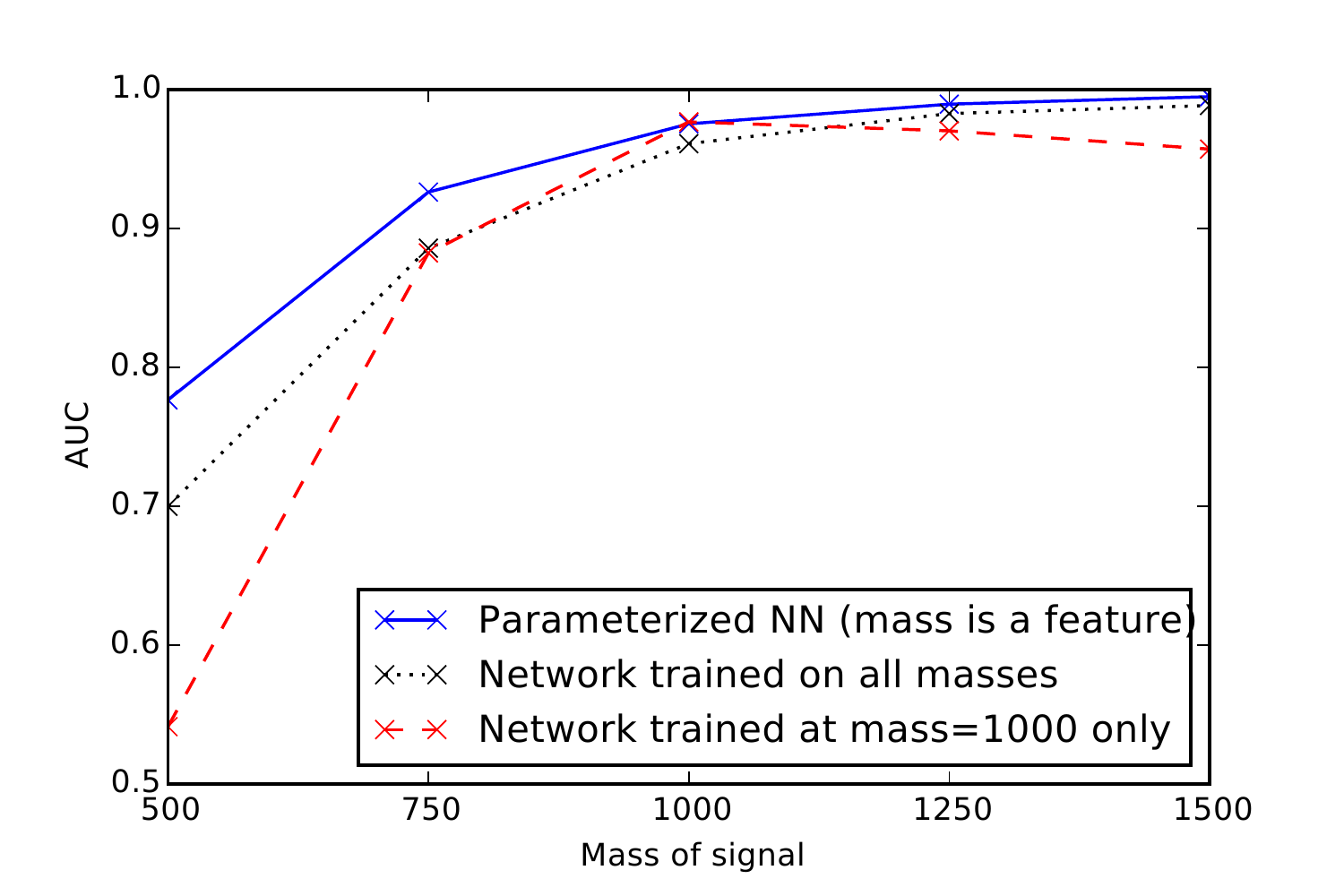}
\caption{ Performance comparison~\cite{Baldi:2016fzo} of signal-background discrimination for a network parameterized by mass (blue), a traditional network trained on all mass values (black), and a traditional network trained only on $m_X=1000$ GeV. As expected, the network trained at a single mass shows decreasing AUC (from ROC curves in Fig~\ref{fig:param_roc}) as the mass deviates from the value in the training sample. The network trained on all masses does not perform optimally at $m_X=1000$ GeV, but the parameterized network performs well at all mass values. The trend of improving AUC versus mass reflects the increasing separation between the signal and background samples with mass, see Fig.~\ref{fig:param_hlvar}.}
\label{fig:param_auc_vmass}
\end{figure}

The high dimensionality of this problem makes it difficult to visually explore the dependence of the neural network output on the parameter $m_{X}$. However, Fig.~\ref{fig:param_auc_vmass} compares the performance of the parameterized network to a single network trained at $m_X=1000$ GeV when applied across the mass range of interest, a common use case. This demonstrates the loss of performance incurred by some traditional approaches and recovered in this approach. Similarly, we see that a single network trained an unlabeled mixture of signal samples from all masses has reduced performance at each mass value tested.

\section{Handling Sets of Four Vectors}

The deep neural network architectures discussed so far have consisted entirely of sequential layers, where each layer is fully-connected to the layer below. However a key advantage of artificial neural network models is the ability to design neural network architectures that reflect properties of the data. These architecture design choices enable us to constrain the class of functions to be considered, or more generally, incorporate an \textit{implicit bias} for some functions over others. Examples include convolutional neural networks, Siamese neural networks, and
various forms of recursive neural networks~\cite{baldi2018inner}.
These architecture designs have been critical to the success of deep learning in computer vision, natural language processing, and bioinformatics. In this section we examine the neural network design choices that can be used to handle \textit{sets} of four vectors in physics.

Neural networks can be designed to have two key properties that are relevant to sets: invariance and equivariance. First, a function implemented by a neural network is \textit{invariant} with respect to an operation if applying that operation to the input does not affect the output. A common example from deep learning is object detection with a convolutional neural network that is invariant to translations of the input image --- the function output could be a single value corresponding to whether an object is present in the image, regardless of whether a translation operation is applied to the image. Second, a function implemented by a neural network is said to be \textit{equivariant} with respect to an operation if applying that operation to the input results in a predictable change in the output. In a convolutional neural network, each convolutional \textit{layer} is equivariant to translations because translating the input leads to a deterministic translation in the output representation. 

For machine learning models that take sets as inputs, it is often desirable to have a model that is invariant or equivariant with respect to \textit{permutation} of the set elements. For example, 3D vision models perform object detection from a set of 3D points on the surface of an object~\cite{qi2017pointnet}, and astronomy models predict the redshift of galaxies from a set of nearby galaxies~\cite{Zaheer:2017,beck2019refined}. Similarly, exotic particle searches in physics involve classifying collision events based on sets of resulting four vectors. The function to be learned should be invariant to the ordering of the set elements. There are at least three ways to try to create supervised neural network models that are invariant to an operation: (1) data-augmentation, (2) canonicalization, and (3) architecture design. We discuss each in turn.

\subsection{Data-Augmentation}

In data-augmentation, the training data is expanded by applying an operation to all training examples. In practice, it is usually more efficient to apply a random operation to the input data at training time. Either way, the network is forced to learn how to be invariant to that operation. For example, in object-detection models it is common to augment the data during training with random translations, rotations, and mirroring operations. On four vectors, ordering and boosting operations~\cite{Butter:2017cot,Pearkes:2017hku} can augment the data. The disadvantage of this approach is that the model must \textit{learn} that the output should be invariant --- this requirement is not enforced by the model. Because this can make the learning problem much more difficult, it is typically used as a last resort when the other methods are not available. 

\subsection{Canonicalization}

The second way to achieve invariance is through \textit{canonicalization} of the input. In this method, the input is always mapped to some canonical element of the group defined by the operation, which enforces invariance without any other constraints on the model. An example is to enforce translational invariance in computer vision by ``centering'' an image at some deterministically-chosen point, or enforcing rotational invariance by rotating the image around that point until it is oriented along some canonical axis (as in Ref.~\cite{baldi2016jet} for jet substructure classification). For sets of four vectors, permutation invariance can be achieved by sorting the particles based on $p_{\textrm{T}}$, as is done in Ref.~\cite{Guest:2016iqz}. A potential disadvantage of this approach, besides having to come up with a good canonicalization scheme, is that the canonicalization procedure can introduce discontinuities in the function to be learned --- a canonicalization that is sensitive to small changes in the inputs is undesirable, and could be worse than no canonicalization at all. 

\subsection{Architecture Design}

The third way to achieve equivariance and invariance in machine learning is through architecture design, where the hypothesis space is constrained to functions that satisfy the condition. For example in convolutional neural network architectures, the convolution layers are equivariant to translations, and together with pooling layers they can be made invariant to translations. 
These architectures have been critical to the success of deep learning in computer vision\cite{urban2016deep}.
Invariance to other input transformations can also be enforced through combinations of weight-sharing and pooling operations. Ideas from Lie group theory can be applied to this problem
\cite{cohen2016group,cohen2019gauge}.
One can trivially define neural network architectures that guarantee invariance to any transformation by defining an ensemble model that applies identical subnetworks to every possible transformation of the input, then pooling the result. Clearly this becomes intractable --- or at least inefficient --- for applications where the number of elements in the group is large or infinite, but there are often simpler approaches.

For input sets, permutation invariance can be achieved by: (1) applying an identical subnetwork to each set element, using shared weights; then (2) pooling the output. The shared weights result in equivariance to permutations of the inputs, since the new outputs will be equivalent to the permutation of the original outputs. The second step achieves invariance, e.g. with max or mean pooling of the possibly-multidimensional outputs. 
Designing neural network architectures that account for data symmetries like permutation invariance is one example of incorporating physics knowledge into the machine learning model, which is discussed more in the next section.

\section{Physics-aware networks}

In applying machine learning to physics problems, one is often presented with the challenge of bringing physics knowledge to bear on the machine learning models \cite{Louppe:2017ipp,Cheng:2017rdo}. This situation can present itself in different forms: choosing of the relevant input and output variables, adding priors or regularization terms in the loss function, or imposing constraints on the neural architectures. Each of these contributes to explicit or implicit model \textit{bias}, which can greatly affect the resulting performance. Often it is difficult to predict how these choices will affect performance, so they are treated as hyperparameters and optimized by trying different variations.
Here we consider two different situations corresponding to physics-informed architecture design and incorporation of physics constraints. 

\subsection{Physics-Informed Architecture Design}

The permutation-invariant models described above are one example of incorporating domain-knowledge into a neural network architecture. We can design neural network architectures that account for additional physics knowledge by taking advantage of other architecture design motifs. These include the local connectivity, weight sharing, and  pooling of convolutional neural networks, but also skip connections~\cite{ronneberger2015u}, gating\cite{hochreiter1997long,cho2014learning}, and attention~\cite{bahdanau2014neural,luong2015effective,vaswani2017attention}. We briefly discuss two other physics-informed neural network architectures applicable to four vectors.

One example of a physics-informed neural network architecture is Ref.~\cite{Louppe:2017ipp}. Decaying particles in the detector typically result in decay products that are hierarchically clustered in space and time (jet substructures). Thus, sets of four vectors often have additional structure that can be exploited. When the clustering hierarchy of each event can be reconstructed, for example using a sequential recombination jet algorithm~\cite{Cacciari:2008gp}, this additional information can be incorporated into the network. \textit{Recursive} neural network architectures can be constructed to match the topology of the jet clustering algorithms, analogous to models from Natural Language Processing that take advantage of sentence parse trees \cite{goller1996learning,socher2011parsing}. The recursive physics neural network architecture is constructed on a \textit{per-event} basis to reflect the tree structure of that event. In addition to the properties of permutation invariance (assuming each node is permutation invariant) and scalability to an arbitrary number of set elements, this model has the additional property of \textit{local connectivity} among related elements in the set, which can lead to better generalization. 


Another example is the Lorentz-Boosted Neural Networks in Ref.~\cite{erdmann2019lorentz}, in which the first hidden layer of the network is interpreted as ``composite particles'' and corresponding ``rest frames,'' and represented as linear combinations of the input four vectors. Each learned composite particle is then boosted into its corresponding rest frame using the non-linear Lorentz transformation. The resulting feature representations are then fed into a neural network, and the entire system is trained using back-propagation. The major advantage of this architecture is that it constrains the representation of the data into a form that is readily interpreted by physicists (i.e. Lorentz-transformed four vectors) and for which physically meaningful features can be extracted such as invariant masses, pseudorapidities, and so forth.

Yet another example is the approach described in  
\cite{fenton2020permutationless}, which uses recursive neural networks, of the form of transformer architectures \cite{vaswani2017attention,devlin2018bert} used in language processing
and tensor attention mechanisms, applied to many-jet event reconstruction in a manner that is invariant to any permutation of the four vectors in the variable-size input set.

\subsection{Incorporating Physics Constraints}


Here we consider the situation where there are physical laws, in the form of exact equations, relating the values of some of the relevant variables. In addition to physics, many fields of science and engineering (e.g., fluid dynamics, hydrology, solid mechanics, chemistry kinetics) have exact, often \textit{analytic}, closed-form constraints, i.e. constraints that can be explicitly written using analytic functions of the system's variables. Examples include translational or rotational invariance, conservation laws, or equations of state. While physically-consistent models should enforce constraints to within machine precision, data-driven algorithms often fail to satisfy well-known constraints that are not explicitly enforced. In particular, while neural networks may provide powerful classification and regression tools for nonlinear systems, they may optimize overall performance while violating these constraints on individual samples. 

Despite the need for physically-informed neural networks for complex physical systems \citep{Reichstein2019,Bergen2019,Karpatne2017a,Willard2020}, enforcing constraints \citep{Marquez-Neila2017} has been limited mostly to physical systems governed by specific equations, such as advection equations \citep{Raissi2017,Bar-Sinai2019,DeBezenac2019}, Reynolds-averaged Navier-Stokes equations \citep{Ling2016,Wu2018}, or quasi-geostrophic equations \citep{Bolton2019}. 
Thus it is necessary to have methods that can enforce analytic constraints in more general settings. Here we describe two general ways for enforcing constraints, first in a soft way, and then in a hard way. 

In general, let us assume that there is a constraint of the form ${\cal C}(x,y,z)=0$ that must be satisfied by the input variables $x$, the output variables $y$, and possibly some auxiliary variables $z$. If $\cal E$ is the error function of the neural network trained on the pairs $(x,y)$, we can enforce the constraints in a soft way by adding a penalty term to the loss function, e.g. using a new loss function of the form ${\cal E}'= {\cal E} + \lambda {\cal C}^2$ where $\lambda$ is an additional hyperparameter controlling the strength of the corresponding regularization (or equivalently log prior) terms. This approach has been used for instance in climate modeling \citep{Karpatne2017,Jia2019,Raissi2020}While this approach can be effective, there is no guarantee that the constraints may not be violated. 

A general way for enforcing constraints in a hard way is described in \cite{beucler2020enforcing}. There are several possible implementation of this idea, but the gist of it is to augment the basic neural architecture with an additional neural network to enforce the constraints. For this, we can first decompose $y$ non-uniquely as $y=(y_1,y_2)$ . Then we introduce a first neural network with adaptive weights that produces an output $y_1'$, trying to predict $y_1$ from $x$. This is followed by a second network which computes $y_2'$ from $x,y_1'$ and $z$, enforcing the constraint $C$ to machine precision. The weights of the second network are fixed and determined by the knowledge of $C$. For instance, the second network can be linear if the constraint $C$ is linear.
We can then combine the two networks into a single overall architecture whose final output is the vector $(y_1',y_2')$. This output always satisfies the constraint $C$ by construction. Furthermore, it can be compared to the target  $(y_1,y_2)$ and the resulting errors can be backpropagated through the combined network, through both the fixed and adjustable weights. As a result of this approach, the constraint $\cal C$ is satisfied at all times, both during and after learning. 

\section{Conclusions}


We have reviewed the advent of deep learning in high-energy physics, first used in classification tasks operating on four-vector features before being applied to tracks, images, graphs, and low-level detector data. 
Even in the case of four-vectors where the number of features is relatively small, deep learning can be used to improve classification performance and incorporate domain knowledge in various forms.
In particular, neural networks can be designed to model a set of related functions using parameterized networks, capture permutation invariance in sets with weight-sharing and pooling, and incorporate additional physics constraints through architecture design or augmented loss functions. 

\clearpage

\bibliographystyle{tepml}
\bibliography{fourv,Constraints,nn,physics}


\end{document}